\documentclass{sig-alternate-10pt}

\usepackage{ifpdf}
\ifpdf
  \usepackage{url}
  \usepackage[pdftex]{hyperref}
  \usepackage[pdftex]{color}
  \DeclareGraphicsExtensions{.pdf, .png, .jpg}
\else
  \usepackage[dvips]{hyperref}
  \usepackage[dvips]{color}
  \renewcommand{\href}[2]{#2}
  \DeclareGraphicsExtensions{.eps, .png, .jpg}
\fi

\usepackage{mathptmx}
\usepackage{graphicx,subfigure}
\usepackage{dcolumn}
\usepackage[super,negative]{nth}        
\usepackage{icomma}                     

\hypersetup{%
  breaklinks,
  baseurl       = http://,%
  pdfborder     = 0 0 0,%
  pdfpagemode   = UseNone,%
  pdfstartpage  = 1,%
  pdfcreator    = \LaTeX{},%
  pdfproducer   = \LaTeX}
  \AtBeginDocument{%
    \hypersetup{%
      pdfauthor = {A. Nonymous},
      pdftitle = {Planet-Scale Human Mobility Measurement}
    }
  }
\def\draftmode{}
\ifx\draftmode\undefined
\newcommand{\comment}[1]{}
\else
\newcommand{\comment}[1]{%
\begin{center}%
\fbox{\parbox[l]{\columnwidth}{\textcolor{red}{#1}}}\\%
\end{center}%
}
\fi

\usepackage[paper=letterpaper, top=1.1in, bottom=1.1in,left=1.05in,right=1.05in]{geometry}

\begin{document}
%

\title{Planet-scale Human Mobility Measurement}

%
%

%


\author{
  Pan Hui$^{\dagger}$,
  Richard Mortier$^{\diamond}$,
  Tristan Henderson$^{\ddagger}$,
  Jon Crowcroft$^{\star}$
  \\
  \\
  \begin{tabular}{ll}
  $^{\dagger}$ Deutsche Telekom Laboratories & $^{\diamond}$ Vipadia Ltd\\
  $^{\ddagger}$ University of St Andrew & $^{\star}$ University of Cambridge\\
  \end{tabular}
}
\maketitle
\begin{abstract}
Research into, and design and construction of mobile systems and algorithms
requires access to large-scale mobility data. Unfortunately,
the wireless and mobile research community lacks such data.
For instance, the largest available human contact traces
contain only $100$ nodes with very sparse connectivity, limited by
experimental logistics. In this paper we pose a \textit{challenge} to the
community: how can we collect mobility data from \emph{billions} of
human participants?  We re-assert the importance of large-scale
datasets in communication network design, and claim that this could  impact
fundamental studies in other academic disciplines. In effect, we argue that
planet-scale mobility measurements can help to save the world. For
example, through understanding large-scale human mobility,
we can track and model and contain the spread of epidemics of various
kinds.
\end{abstract}

\section{Introduction}
\vspace{-2mm}
Human mobility traces are critically important to many disciplines in
addition to computer networking, ranging from
epidemiology~\cite{colizza-2007-5} to urban
planning~\cite{strano07centrality}.  Unfortunately, existing traces of
human mobility are flawed: using traditional social science methods to
collect data has proven difficult~\cite{smallworld}, and traces
collected using technology methods have suffered from a variety of
limitations.  These include small size (the largest is 100
nodes~\cite{psn-mobihoc}), short duration (the longest is 9
months~\cite{realityMining}), and high locality (many of the scenarios
are limited to campus and conference environments~\cite{psn-tmc07}).
These datasets may not be enough for large mobile system
evaluations, and are definitely insufficient for epidemiology, where
planet-wide measurements are needed to track the spread of disease.

As members of the networking community, we have both the tools and methods (e.g.,~hardware
and software knowledge) to conduct large-scale data collection.
Furthermore, our contributions will not only benefit the wireless and
mobile networking research communities, but will impact fundamental
research in other areas allowing more features about human behaviour
to be uncovered.  We believe that the situation is analogous to that
of complex networks research, which has flourished since 1989 when the
first large datasets from the Internet (and subsequently the World
Wide Web) became available~\cite{albert-2002-74}.  To achieve similar
improvements in mobile networking and other related fields, relevant
large-scale datasets must be made available.

In this paper we challenge the community to collect large-scale human
mobility traces.  We highlight some of the issues in the hope that the
community can help find good solutions.  In the meantime, we propose
some solutions intended to form the basis of initial efforts; the main
aim is to raise these issues to gain community support to meet this
challenge and make the topic \emph{hot} in the networking community.
\vspace{-2mm}

\section{Why are large-scale human mobility traces important?}
\vspace{-2mm}
As mentioned above, large-scale datasets are useful for many
aspects of research. In this paper we focus only on two aspects:
system design and validation, and epidemiological studies.

\subsection{System design and validation}
\vspace{-1mm}
After its first use in the evaluation of Dynamic Source
Routing~\cite{Johnson96dynamicsource}, the random waypoint model
became the \emph{de facto} standard mobility model in the mobile
networking community. For example, of the ten papers in ACM MobiHoc 2002
which considered node mobility, nine used the random waypoint
model~\cite{yoon:random}.  This trend has changed dramatically over
recent years after the introduction of real mobility traces for
evaluation: of the 10 papers considering node mobility in MobiHoc 2008,
7 used real mobility traces for evaluation.

The community has realised that unrealistic models are harmful for
scientific research.  Although real traces may suffer from limited
numbers of participants, coarse granularity, and short experimental
duration, they at least reflect \emph{some} aspects of real life.
Thanks to the popularity of Online Social Networks (OSNs), we can now
gather large-scale data about the topology and membership information
of millions of OSN users and use these to study aspects of the social
networks~\cite{mislove-2007-socialnetworks,lewis:facebook}. But where
is the large-scale dataset for evaluating, for instance, inter-city
ad-hoc communication using mobile computing?  Or even a single
city-wide mobile communication system (e.g.,~a delay-tolerant network, or
city-wide gaming)?  We have very few empirical hints for this.
Without the help of real data, we cannot even know whether this kind
of system is possible.  Even if we extrapolate large-scale
mobility traces from small-scale traces, the problem of validating the
extrapolation remains.

Instead of using mobility traces directly to run trace-driven
simulations, a possible approach is to extract characteristics from
the data and build more realistic mobility models.  Much work has been
done in modeling human mobility for mobile ad hoc network
simulation~\cite{Camp02asurvey}.  Researchers have proposed more
realistic models by incorporating obstacles~\cite{jardosh03mobility},
social information~\cite{musolesi:models}, and clustering features
observed in mobility datasets~\cite{comsnets09piorkowski}.  Analysis
of real traces has demonstrated power-law inter-contact time
distributions with cut
off\\~\cite{psn-tmc07,Karagiannis07mobile}, levy-flight
patterns consisting of lots of small moves followed by long
jumps~\cite{rhee:levymobility}, heterogenous
centralities~\cite{freeman1977} (i.e.,~popularity) and clustering
structure~\cite{psn-mobihoc}.  But again, these results are from
small-scale datasets and are limited to specific scenarios with
limited time durations.  Some researchers have extrapolated from these
by assuming, for instance, that the way people move in a city is
correlated to the centrality distribution of the city
graph~\cite{strano07centrality}, but this has yet to be verified
empirically.  Gonzalez \emph{et al.}~\cite{gonzalezNature08} extracted
coarse-grained levy-walk properties from large-scale mobile phone
usage.  The limitation is that the dataset is from cellular
basestation, which only log when mobile users make a call or send an
text message.  This is very coarse in both geographical and temporal
granularity.  People may argue that human behavior should be
scale-free in different dimensions, but we need data for further
verification. Moreover, since the the data from this study have not
been released, it is impossible to verify or build on their findings.

We need large-scale human mobility datasets with better space and time
granularity to verify the properties we mention above.  Following
analogous progress in related fields, it seems likely that we
will uncover many more features from such data which will help us to
build good models.  We believe that this is crucially important for
the mobile computing community.
\vspace{-1mm}
\subsection{Epidemiological studies}
\vspace{-1mm}
Moving beyond social science, the communication network community has
also aided research in many other academic disciplines.  For instance,
our methodology and data make the modeling of human dynamics
possible~\cite{vazquez-2006-73}, and more significantly, our data made
possible the development of the field of complex network
research~\cite{albert-1999-401}.  Large-scale mobile data can further
enable the study of epidemic disease spreading. 
The current state-of-the-art in
epidemic modeling uses data from the International Air Transport
Association (IATA) commercial airline traffic database
to determine travel between airports and to provide coarse-grained estimates of
global spreading patterns~\cite{colizza-2007-5}, as well as data of
transportation and commuting patterns in urban areas, which can be
used to model a metapopulation mechanism of
spreading~\cite{colizza-2008-251}.  Researchers cannot develop more
microscopic models of epidemic spreading because of the lack of
large-scale fine-grained empirical data.

To take a topical example, consider the current swine flu outbreak.
Scientists have urged governments to map the spread of
swine flu more accurately in order to predict the number of people who
may die from it~\cite{garske:severity}.
Current predictions indicate that one in 200 people who get swine flu badly enough to
need medical help could go on to die, but given that vaccines may not be ready until
later than hoped, accurate predictions are crucial.
available.
Any estimates about swine flu are subject to a wide margin
of error, not least because not everyone who catches it develops
symptoms. More accurate mapping of the spread of the virus must be
carried out if it is to be effectively managed. Monitoring doctors and
hospitals is insufficient since not everyone who is
infected with swine flu will become ill enough to report their case to
a doctor.
\begin{figure}[t]
\begin{center}
\includegraphics[width=7cm]{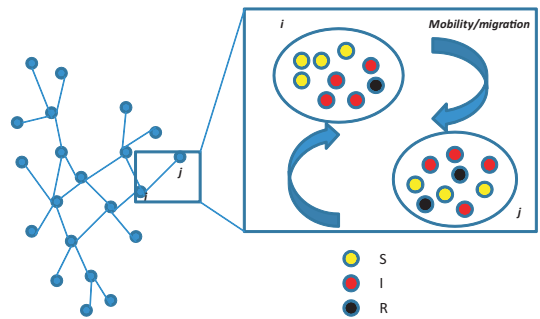}\vspace{-1mm}
\caption{\label{migration} Metapopulation model composed of a network of subpopulation connected by mobility.}
\vspace{-3mm}
\end{center}
\vspace{-4mm}
\end{figure}

Figure~\ref{migration} shows the process of the spreading of epidemics
by the mobility of human from a subpopulation (e.g., a city) to
another subpopulation. When a susceptible individual (S) is in contact
with a infectious individual (either symptomatic or asymptomatic), it
will be infected with a certain rate and enter the latent class. When
the latent period ends, the individuals become infectious (i.e., able
to transmit the infection). After the infectious period, all
infectious individuals enter the recovered class. If an infectious
individual moves from to another city, the subpopulation in the new
city will also be infected. Using the IATA data, scientists can
roughly model the migration of population across countries. But we need
much better granularity of data, instead of assuming a homogeneous mixing
in each subpopulation (city).
\begin{figure}[t]
\begin{center}
\includegraphics[width=8cm]{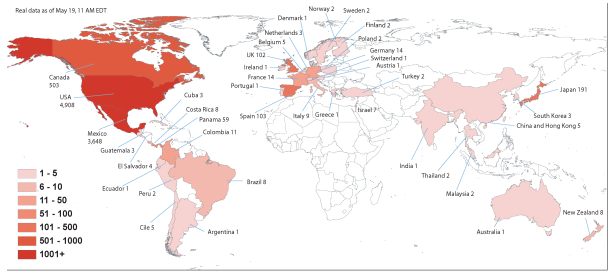}\vspace{-1mm}
\caption{\label{viewGlobalSwine} Confirmed number of swine flu cases worldwide on May 19 (from GLEaMviz.org)}
\vspace{-3mm}
\end{center}
\vspace{-4mm}
\end{figure}

Figure~\ref{viewGlobalSwine} shows the confirmed number of swine flu
cases world wide on May 19, 2009. It first started in Mexico and then
spread to other countries by human mobility. We can see from the
figure that the most worst countries beside Mexico are nearby
countries such as the USA and Canada. Spain was the worst in Europe
because there are in general a lot of connections between Spain and
Mexico, but we need better data to build models to predict such
behaviour.

Mobile computing can help to fight epidemics in at least two ways:

\textbf{Case 1:} If we can track real-time or nearly real-time human
health status, we can provide advice and precautions for each users,
accurately estimate the number of asymptomatic infectious individuals,
predict the spreading process, identify the hotspots of the
pandemic, and effectively isolate the infectious victims. This may be
possible by using a personalised epidemic software. Users can self
identify their health status (e.g., cough, cold) and embedd this
status in a Bluetooth service. Users periodically run Bluetooth service
discovery and log the devices discovered, the health
status of each encountered user, and if possible also their geographical
locations. Users can upload their log files to the server, which
analyses results and provide effective feedback.

\textbf{Case 2:} If we do not have the health status of each users but
only the contact log and the geographical location of certain
encounters, we can understand the mixing properties of each
subpopulation, model contact and mobility processes, and identify the
social hotspots. With this understanding, we can accurately predict
and emulate the spreading of diseases.
\vspace{-1mm}
\section{Challenges in collecting data}
\vspace{-2mm}
\subsection{High experimental cost}
In general it is expensive to conduct large-scale mobility
experiments.  Costs include equipment, software, human resources, and
generating incentives for people to participate.  For example, for the
iMote experiments carried out by the Haggle
Project~\cite{psn-mobihoc}, the cost of iMotes, packages, batteries,
participation incentives, and the human resources spent on
assembling and distributing devices, and monitoring the experiments.
add up to \$12,000 (including development) for a
small-scale experiment with just 50 participants. This is clearly not scalable to
experiments involving billions of people.
\vspace{-1mm}
\subsection{Privacy and government regulations}
\vspace{-1mm}
The law in many jurisdictions strictly regulates privacy and thus data collection,
making large-scale data collection even more
challenging~\cite{henderson:hotplanet}.  Before data collection can
begin, the consensus of participants is required, substantially
increasing the administrative burden.  Further, telephone operators
are restricted in what customer data they can store, for how long, and
for what purpose, and the dissemination of such data is even more
tightly controlled.  This dramatically increases the difficulty of
obtaining data from operators, which otherwise is a good way to reduce
collection cost and increase dataset size.

\vspace{-1mm}
\subsection{Lack of motivating applications}
\vspace{-1mm}
We can see from the discussion above that it is not scalable to give
out hardware for large-scale experiments.  Instead we must rely on
useful or interesting software applications to motivate participation
of users that already own their own hardware.  For example, there are
many applications developed for iPhones but no key application exists
that enables large-scale data collection.  An application able to
scale up to millions of users while collecting data would be
incredibly valuable to the research community (as well as
economically!).  Equal value might be obtained through many
applications with smaller (but still large) user communities: it is
not a strict requirement that such a large dataset consist of a single
community, and indeed, it might be valuable in avoiding bias if the
overall billion-sized dataset were composed of numerous smaller
(multi-million sized) components.
\vspace{-1mm}
\subsection{Lack of business models}
\vspace{-1mm}
To motivate a large amount of participation, we may need good business
models. Such business models can motivate operators to share their
data, and users to participate in experiments.
If all parties --- the operators, the
users and the researchers --- can benefit from participating in a
system, it is more likely to succeed.
\vspace{-1mm}
\subsection{Lack of organisation}
\vspace{-1mm}
CAIDA (\url{caida.org}) exists to aid Internet
traffic data collection, but there is no such organisation or group for
data collection in mobile or wireless networks.
The closest is CRAWDAD (\url{crawdad.org}), but that was established only to \emph{archive}
wireless data and, though it has performed this role well, it
does not currently coordinate or lead data collection.  An
organisation for initiating, motivating, and coordinating
mobile data collection would be extremely valuable.  If such an
organisation cannot be founded then, given the distributed and
large-scale nature of the problem, crowd-sourcing might be utilised to
achieve the same goal.
\vspace{-2mm}
\section{What can we do?}
\vspace{-2mm}
It is impractical to provide experimental devices to billions of
participants. Our strategy is to develop novel application software
allowing us to utilize crowd-sourcing.  It is also impossible to
collect data from billions of people while relying on one group alone:
we need collaborative support from the joint force of the research and
industrial communities to achieve active participation of sufficient
individual users.  The key problem is to motivate participation of the
community and users by providing mutual benefit.
\vspace{-1mm}
\subsection{New communication and networking applications }
\vspace{-1mm}
Novel and useful communication and networking applications can be one
efficient way to motivate participation. For example the company
SenseNetworks \\(\url{www.sensenetworks.com}) provide a
innovative mobile application for real-time nightlife discovery and
social navigation, answering the question, ``Where is everybody going
right now?''  They found that this application attracted around
100,000 users in North America.  Unfortunately, as with other
companies, the data are not available to the public but it seems that
developing useful applications might be a viable way to collect
large-scale datasets for research purpose.

Another example from the research community are applications designed
to encourage users to share their mobile phone
devices~\cite{liu:xshare} or calling minutes and text
messages~\cite{psn-shair}. Such applications provide an incentive for
usage and so could be used to motivate participation in experiments.
\vspace{-1mm}
\subsection{A common research platform for mobility and social network study}
\vspace{-1mm}
Currently there are several research groups involved in human
mobility
measurements~\cite{psn-tmc07,rhee:levymobility,Karagiannis07mobile,Vikram06Mobicom,bigwood08Wimob},
and we expect more researchers will move into the area in the near
future. Social network research has also recently become a popular
research area, and is often integrated with mobility
research. In order to motivate the researchers to create a
crowd-sourcing effect, we propose the development of an open platform for
social network and mobility experiment. Researchers can create their
own online social networks for their research projects by defining the
fields of users' profiles according to the need of their experiment,
e.g.,~name, email addresses, and Bluetooth ID. Separate projects can
have different users, but the platform itself will merge the database
from all projects.  When a new project starts the central server
informs all users about this project and invites them to
participate. The user interface and format for each project are
similar, and projects can be merged on the platform. The different is
that each project has a database, and manages its own data
independently. This will save a lot of effort and administrative
hassle when collecting and interpreting data, and conducting
experiments.
%
\vspace{-1mm}
\subsection{A social proximity application}
\vspace{-1mm}
Isolation is usually a problem in metropolitan cities. Mobile devices
can help to detect the devices in proximity and help people to notice
the ``familiar strangers'' around them.

Mobile phones can sense the people we meet everyday within the radio
range and also detect the duration of the proximity. Here we suggest a
platform including both software running on the mobile client and a
web based application, allowing the users to build up a social network
based on the proximity information detected. Mobile users can create a
profile page on the web server by register its Bluetooth ID.  The
profile page can be similar to a Facebook page, but having additional
features which can allow the user to preview statistics about the
people he met in any period, and propose related strategies for
subsequent encounters. The user can request addition of a particular
owner of a Bluetooth ID to his friend list as on Facebook. We believe
this opens a completely new way of socialising.

For example a user could use his mobile phone to detect the Bluetooth ID
of someone whom he sees on the subway everyday, but to whom he is too
scared to talk.
This could enable him to initiate contact, while leaving the other
party
in control of any communication. This application
scenario may seem socially unlikely in the Western world but it is a
common pattern in Asia.
But note
that a single Asian population, however large, is also
unrepresentative: many suitable applications,
encouraging participation from different continents, countries
and cultures, may be necessary.

%
\vspace{-1mm}
\subsection{Request data from the operators}
\vspace{-1mm}
We have two ways to request data from the operators: either access to
anonymised data e.g.,~via collaborative research projects; or full
access to data as a commercial partner, e.g.,~by providing commercial
value to the operator through data analysis. An example of the former
is the access of the Google metropolitan Wi-Fi
dataset~\cite{afanasyev08googlewifi}. This might be possible if the data can
help to improve their services or provide them better revenues, e.g.,
if understanding human mobility can help in Wi-Fi hotspot deployment
and placement.
For the latter approach, one good
example is applications like Qiro (\url{qiro.net}) or
SenseNetworks, both of which use collaboration with operators to
access location information to provide additional services to the
users. Qiro uses information from T-Mobile, E-Plus, Vodafone and O2 to
help users to locate nearby friends, and facilities such as bicycle
rental.
%
%
\vspace{-1mm}
\subsection{Collaboration with local government and media}
\vspace{-1mm}
Local governments are powerful entities for assisting with data
collection. They can help to push applications into reality. Some
governments seek to develop infrastructure and facilities to improve
the people's life in the metropolitan area. By collaborating with
these governments, we can quickly access the resources and deploy the facilities. The local media can be also a
good way to gather mobility information as they are often interested
in new technologies, wanting to use them in future campaign
activities. For example, to market the movie
\textit{Artificial Intelligence}, an augmented reality
game based on the movie, called \textit{Beasts}, was created. The game
was conceived as an
elaborate murder mystery played out across hundreds of websites, email
messages, faxes, fake advertisements, and voicemail messages, and
involved over three million active
participants. 
Collaborating in such activities can gain us
datasets of millions of people. The UK government for the swine flu case can also be a good collaborator for the data collection.
\vspace{-1mm}
\subsection{New sources of data}
\vspace{-1mm}
The popularity of Web 2.0 and user-generated content means that there
may well be more available human mobility datasets on the Internet, if
we know where to look.
For example,
Piorkowski was able to extract
125,000 short-term mobility traces gathered from a publicly available
web-based repository of GPS tracks~\cite{Piorkowski:hotplanet} - the Nokia Sports Tracker service, which covers mobility of many urban areas. Another example is photo-sharing sites like Flickr. Photo-sharing sites on the Internet contain billions of publicly accessible
images taken virtually everywhere on earth, which are annotated with various forms of information including geolocation, time, photographer, and a wide variety of textual tags. Researchers have been able to analyse a global collection of geo-referenced photographs, and evaluate them on nearly 35 million images from Flickr~\cite{crandall09:flickr}.

We believe in order to achieve the goal of planet-scale mobility
measurement, we need to be more creative in collecting and merging
information from different sources, sensing methods, and collaborating with different organisations.
%
%
\vspace{-2mm}
\section{Conclusion}
\vspace{-2mm}
In this paper we challenge the networking community to collect
planet-scale human mobility traces. We explained why large-scale
mobility datasets are important for networking research, and how they
could impact fundamental researches in many other academic
disciplines. We identified the challenges and difficulties, and
further proposed potential methods to achieve this goal.

We in no way claim that we have the ideal strategies for collecting
and managing such datasets: we would go so far as to say that this is
an impossible mission for a single research group. Our intent with
this paper is to draw the attention of the community to this problem,
enabling the collective intelligence of the whole community to be
brought to bear on these crucial problems.

With these kind of datasets, we believe that we will completely change
the understanding of human dynamics, potentially opening many new
fields of academic study, as the availability of Internet and WWW web
data allowed the study of complex networks and systems to flourish,
further impacting the understanding of biological structures,
e.g.,~DNA and proteins.  We urge the community to address these
challenges to make this possible, and in doing so perhaps we can help
to save the world from epidemics like SARS and swine flu.

\bibliographystyle{abbrv}
\bibliography{million}
\end{document}